 \documentclass[smallabstract,smallcaptions]{dccpaper}

\usepackage{epsfig}
\usepackage{citesort}
\usepackage{amsmath}
\usepackage{amssymb}
\usepackage{color}
\usepackage{url}
\usepackage{multirow}
\usepackage[caption=false]{subfig}
\usepackage{graphicx}

\newlength{\figurewidth}
\newlength{\smallfigurewidth}

\DeclareCaptionSubType[alph]{figure}
\makeatletter
\renewcommand\p@subfigure{\thefigure~}
\makeatother

\begin{document}

\title
{\large
\textbf{JQF: Optimal JPEG Quantization Table Fusion by Simulated Annealing on Texture Images and Predicting Textures
}
}

\author{%
Chen-Hsiu Huang and Ja-Ling Wu
\\[0.5em]
{\small\begin{minipage}{\linewidth}\begin{center}
\begin{tabular}{c}
Department of Computer Science and Information Engineering, \\
National Taiwan University \\
No. 1, Sec. 4, Roosevelt Rd., Taipei City 106, Taiwan (R.O.C.) \\
\url{{chenhsiu48,wjl}@cmlab.csie.ntu.edu.tw} 
\end{tabular}
\end{center}\end{minipage}}
}

\maketitle
\thispagestyle{empty}

\begin{abstract}
JPEG has been a widely used lossy image compression codec for nearly three decades. The JPEG standard allows to use customized quantization table; however, it's still a challenging problem to find an optimal quantization table within acceptable computational cost. This work tries to solve the dilemma of balancing between computational cost and image specific optimality by introducing a new concept of texture mosaic images. Instead of optimizing a single image or a collection of representative images, the simulated annealing technique is applied to texture mosaic images to search for an optimal quantization table for each texture category. We use pre-trained VGG-16 CNN model to learn those texture features and predict the new image's texture distribution, then fuse optimal texture tables to come out with an image specific optimal quantization table. On the Kodak dataset with the quality setting $Q=95$, our experiment shows a size reduction of 23.5\% over the JPEG standard table with a slightly 0.35\% FSIM decrease, which is visually unperceivable. The proposed JQF method achieves per image optimality for JPEG encoding with less than one second additional timing cost. The online demo is available at \url{https://matthorn.s3.amazonaws.com/JQF/qtbl_vis.html}.
\end{abstract}

\Section{Introduction}

\begin{figure}[!ht]
\centering
\subfloat[]{{\includegraphics[width=5cm]{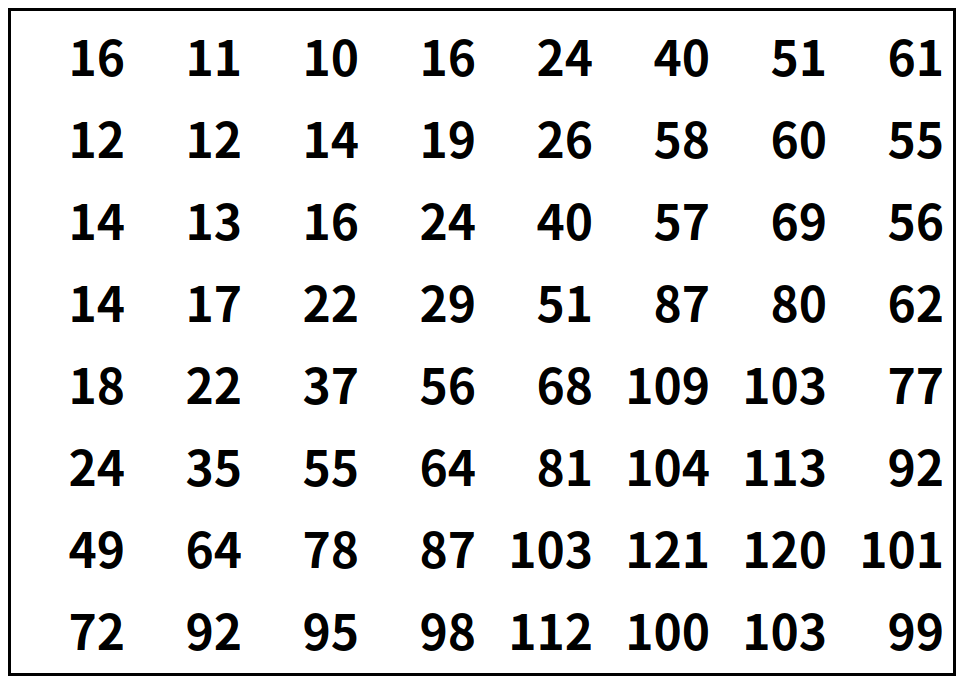} \label{fig:std_luma}} }%
\subfloat[]{{\includegraphics[width=5cm]{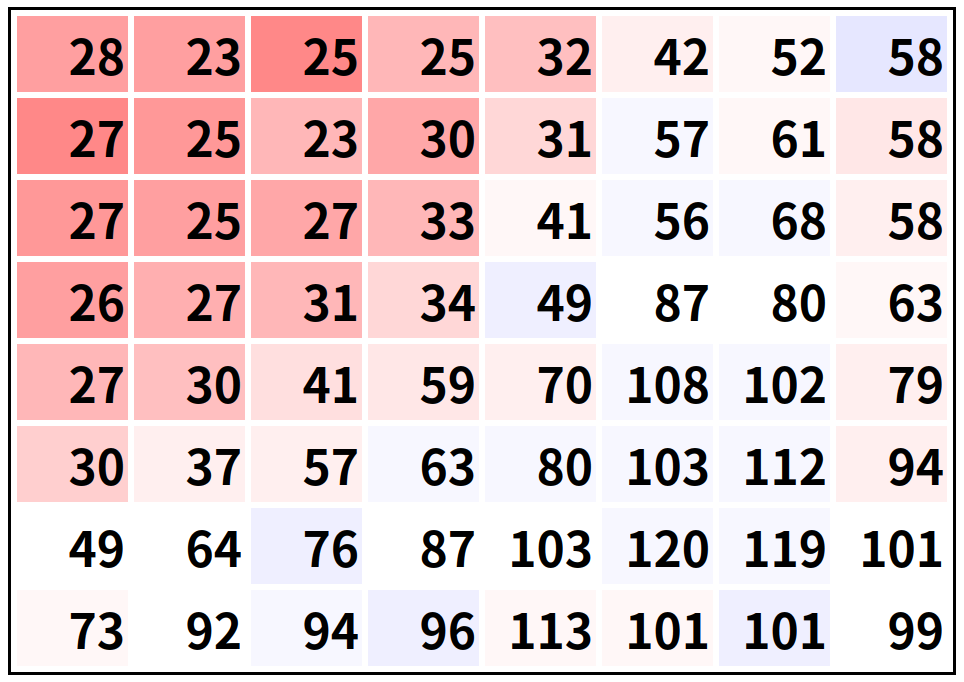} } \label{fig:fuse_lh} }%
\caption{Examples of JPEG luminance quantization table: (a) the default standard table, and (b) the fused customized table for the lighthouse image} %
\label{fig:qtable-ex}%
\end{figure}

JPEG is a commonly used lossy compression standard for digital images, developed by the Joint Photographic Experts Group \cite{wallace1992jpeg} in 1992. Although the JPEG Still Picture Compression Standard has been introduced for nearly three decades, JPEG remains the most frequently used image format, whether in the Internet content sharing \cite{hudson2018jpeg} or produced by various digital image capture devices. JPEG divides the image into 8x8 blocks, using Discrete Cosine Transform (DCT) to shift the pixels from spatial domain to frequency domain for better coding efficiency. Because the human visual system (HVS) is more sensitive to low-frequency components and less perceivable on high-frequency components, the transformed DCT coefficients are rearranged in zigzag order for reflecting the spectral importance. Then a quantization table with values in different magnitude is used to quantize DCT coefficients in corresponding spectrum position, resulting in reduced coefficient values and sparse DCT block, which is more beneficial to variable length coding and run-length encoding (RLE). Figure \ref{fig:std_luma} shows the default luminance quantization table provided in the JPEG standard. It is easy to observe that quantization values are generally increasingly arranged in the zigzag scanning order. 

Since JPEG is lossy, the compression rate can be adjusted, allowing a selectable tradeoff between storage size and image quality. \label{sec_scale} The JPEG standard library \cite{independent2016libjpeg} includes a quality metric $Q$, ranging from 1 to 100, to scale values in the quantization table to control the reduction of DCT coefficients. The \texttt{libjpeg} reference implementation demonstrates how to calculate the scaling factor $S_f$ and scale quantization values to target quality: 

\begin{equation} \label{eq:scale_factor}
S_f = \begin{cases}
50 / Q \quad & 1 \le Q < 50 \\
2 - Q/50 \quad & 50 \le Q \le 100 \\
\end{cases}
\end{equation}

\begin{equation} \label{eq:scale_quant}
T_Q(i) =  \max(\min(T_S(i) \times S_f, 255), 1)\quad,\quad \text{for } i=1 \dots 63
\end{equation}

where $T_S$ is the standard quantization table and $T_Q$ is the scaled table at quality $Q$. 
Although a smaller quality metric achieves better image size reduction, it may introduce visual artifacts such as blocking effect and ringing effect if we carefully observe image pixels under a magnifier. Therefore, the default JPEG quality metric in various applications is usually set to higher values, say at least 75 or above. Except for the standard table, the JPEG standard also allows users to use customized quantization tables. However, the JPEG quantization table's selection remains a challenging and un-optimized problem due to the numerous solution space and lack of reliable quality measurement that accurately model the HVS. As a result, the modern image applications and digital camera image processors tend to compress JPEG images with minimal quantization values to preserve the quality. 

In this paper, we try to solve the dilemma of balancing between computational cost and image specific optimality by introducing a new concept of texture mosaic images. We use the RAISE dataset \cite{dang2015raise} as our training database and cross-validate on the Kodak dataset \cite{kodakcd}. We crop the training images into patches and further apply unsupervised clustering to categorize different texture types, then stitch those texture patches to form the texture mosaic images. The simulated annealing technique is used on those texture mosaic images to search for an optimal quantization table for each different texture category. The convolutional neural network (CNN) model is applied to the texture clustering result to learn the generic representation of texture features, and used to predict the testing image's texture distribution. Finally, based on each image's texture characteristics, the texture relevant optimal quantization tables are fused to come out with an image specific optimal quantization table. On the RAISE testing set and the Kodak dataset, our experiments show a double-digit percentage size reduction compared to the JPEG standard table with a slight decrease of FSIM score, which is visually unperceivable under high quality metric setting. 

\Section{Related Works} \label{sec_related}

\SubSection{Simulated Annealing}

The stochastic optimization process known as simulated annealing has been applied to find vector quantization parameters. The works from Monro and Sherlock \cite{monro1993optimum,sherlock1994model} were the first attempts to use simulated annealing on determining quantization tables for DCT coding. To locate an optimized quantization table, we applied the simulated annealing to all 64 quantization values with the cost function composed of RMSE error and a selected target compression ratio. Their optimization process searches optimal tables on selected images with minimal RMSE error while keeps the compression ratio close to the chosen target. Around a single digit percentage of error improvement is reported comparing to the standard JPEG table. Both Monro's and Sherlock's works indicated the following two things: 1) High frequency components are more critical than the assumption made in the JPEG standard table; 2) RMSE can only be used as a power based measure for signal fidelity, but shown to be an inferior metric to approximate subjective image quality. 

Until the early 2000s, new objective FR-IQA methods like SSIM \cite{wang2004image} and FSIM \cite{zhang2011fsim} were proposed and shown to be statistically closer to HVS. Jiang et al. \cite{jiang2011jpeg} utilize SSIM as the quality metric to evaluate distortion in the compressed images during the simulated annealing process. In their work, a multi-objective optimization equation is proposed to minimize bitrate while maximizing SSIM. To solve the equation, they estimate the Pareto optimal point for finding an optimal quantization table. There are no other feasible points to have both lower bitrate and higher SSIM index. Their best annealing technique achieves 11.68\% size reduction over the JPEG standard table while slightly decreases the SSIM index by 0.11\%. However, since Pareto optimal point differs in every image, the multi-objective optimization framework only proves useful on a per-image basis, not on a set of evaluation images. 

On top of Jiang's work, Hopkins et al. \cite{hopkins2018simulated} adopt FSIM as the quality metric and revise the annealing process to focus on compression maximization with a temperature function that rewards lower error. A set of 4,000 images was selected from RAISE \cite{dang2015raise} dataset as a training set to run four groups of 400 separate annealing processes in parallel at quality metrics 35, 50, 75, and 95. With the four global optimized quantization tables at different quality metrics, Hopkins' work reduces the compressed size by around 20\% over the JPEG standard table. It claims to improve FSIM error by 10\% on the evaluation set. The corpus of 4,000 training images looks like a pretty good proxy to the universal pictures, but still not custom-tailored per image. 

Another work from Google's JPEG encoder Guetzli \cite{alakuijala2017guetzli} aims to produce visually indistinguishable images at a lower bit-rate with Butteraugli \cite{butteraugli2016}, Google's perceptual distance metric. By using a close-loop optimizer, Guetzli optimizes global quantization tables and selectively zero out specific DCT coefficients in each block. Compared to Hopkins' global optimized quantization table, Guetzli's per image optimization strategy achieves a 29-45\% data size reduction. However, the most size reduction improvement of Guetzli comes from identifying DCT coefficients to zero out, not from optimizing the quantization table. Google's Guetzli provides both per image customized optimization and better size-reduction, but is extremely slow (up to 30 minutes on a high-resolution image) and considered not for practical use. 

\Section{Proposed Method} \label{sec_framework}

\begin{figure}[!ht]
\centering
\subfloat[]{{\includegraphics[width=13cm]{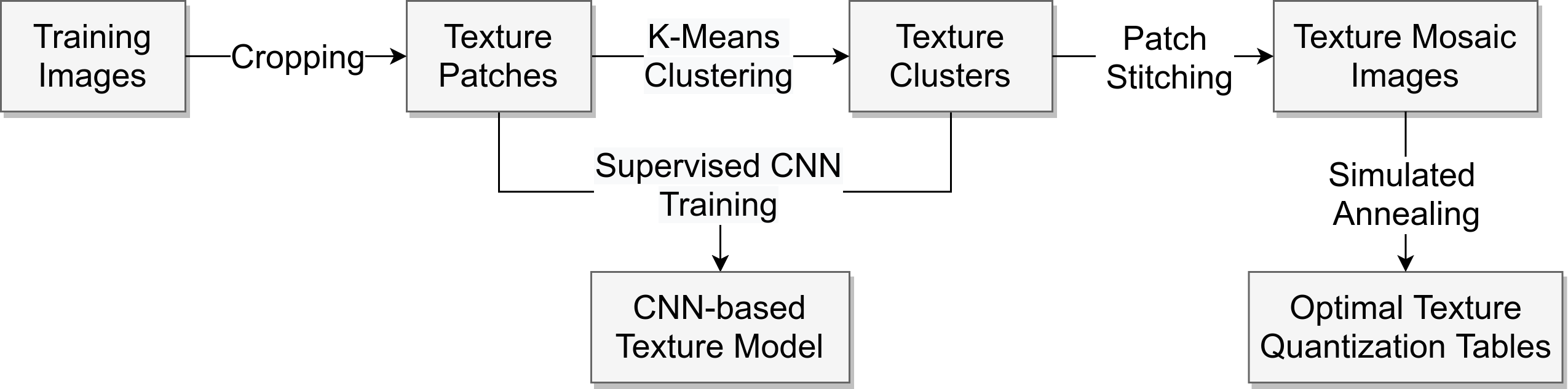} \label{flow_1}}}%
\\
\subfloat[]{{\includegraphics[width=13cm]{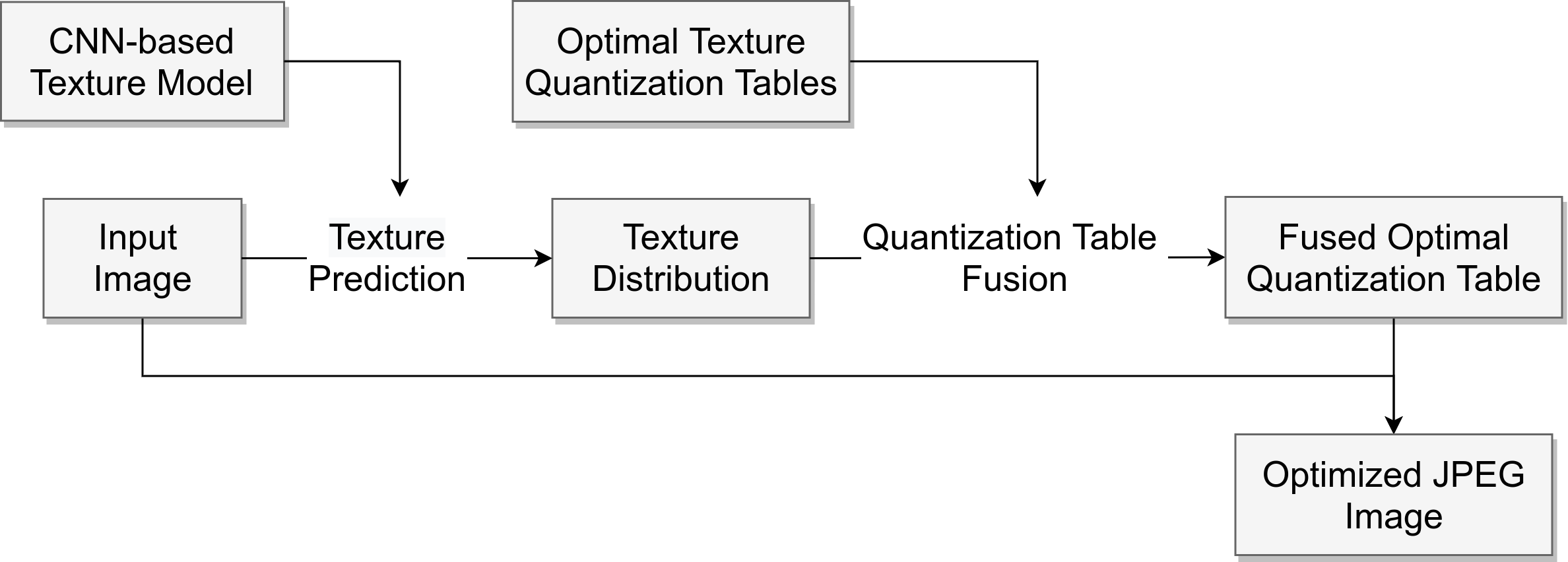} } \label{flow_2}}%
\caption{The workflows of the proposed JQF method. (a) The texture training and mosaic image annealing flow. (b) The texture prediction and quantization table fusion flow. }%
\label{fig:workflow}%
\end{figure}

The proposed JPEG Quantization Table Fusion (JQF) method contains two workflows. The training workflow is shown in Figure \ref{flow_1} served as a series of offline procedures to collect and cluster texture patches from the training image dataset, then optimize the texture quantization tables. In the prediction workflow shown in Figure \ref{flow_2}, we predict the texture distribution of the input image by the texture CNN model, then aggregate a custom-tailored quantization table for the input image. 

\SubSection{Texture Patches Clustering} \label{sec_cluster}

We crop the training images into $64\times64$ patches in a non-overlapping manner and perform unsupervised image clustering methods on textures. Typically the image clustering problem can be handled with two steps: 1) finding the appropriate image visual features and 2) training a classifier that minimizes the class assignments \cite{caron2018deep}. In recent years, the pre-trained CNN models on ImageNet \cite{krizhevsky2012imagenet} have become the building blocks in many computer vision applications. In this context, the last activation maps after layers of ConvNets, called bottleneck features, is a good choice of visual features to represent raw pixels in a vector space of fixed dimensionality. 

\begin{figure}[!ht]
\centering
\subfloat[]{{\includegraphics[width=5cm]{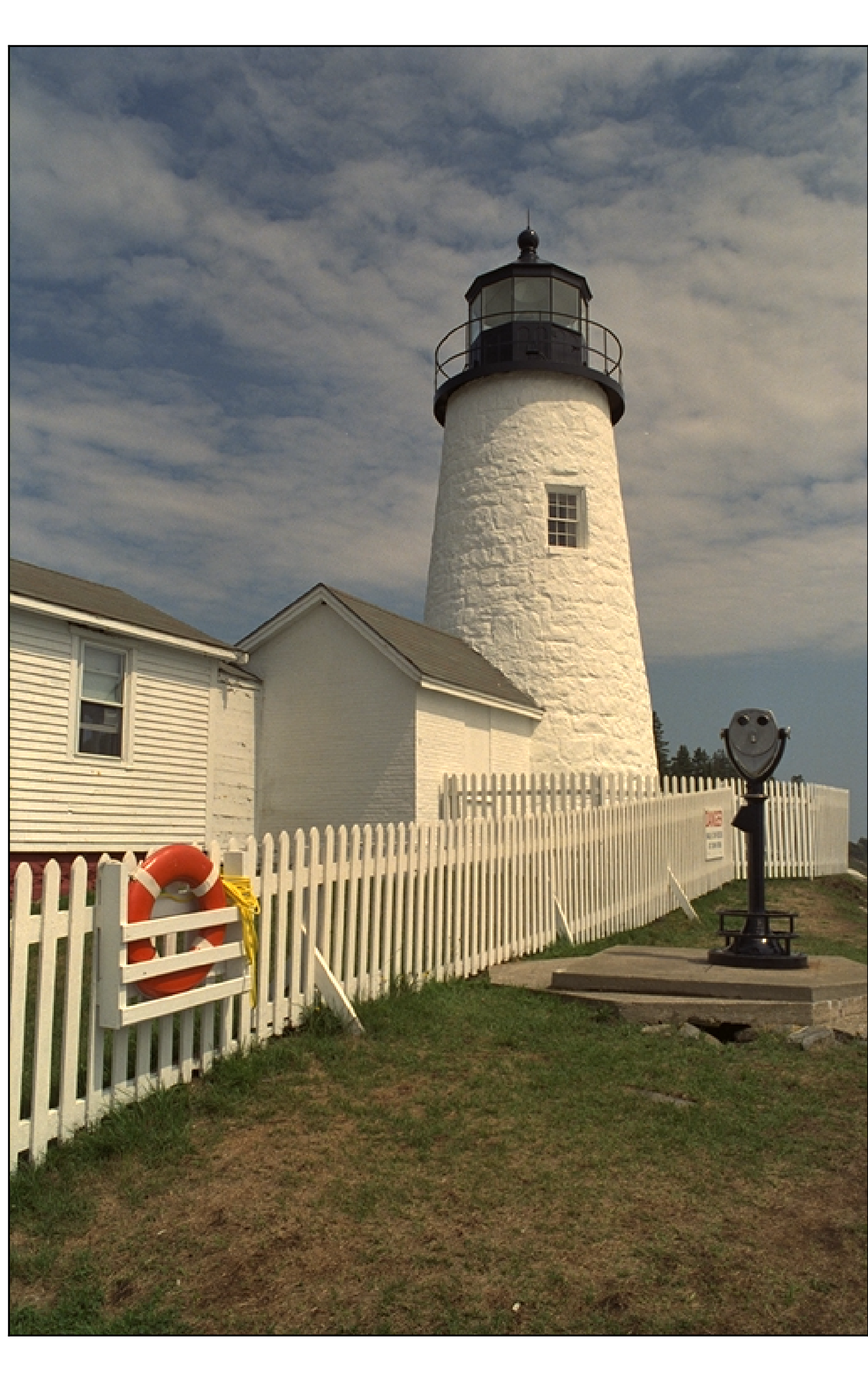} \label{lh_src}} }%
\subfloat[]{{\includegraphics[width=8cm]{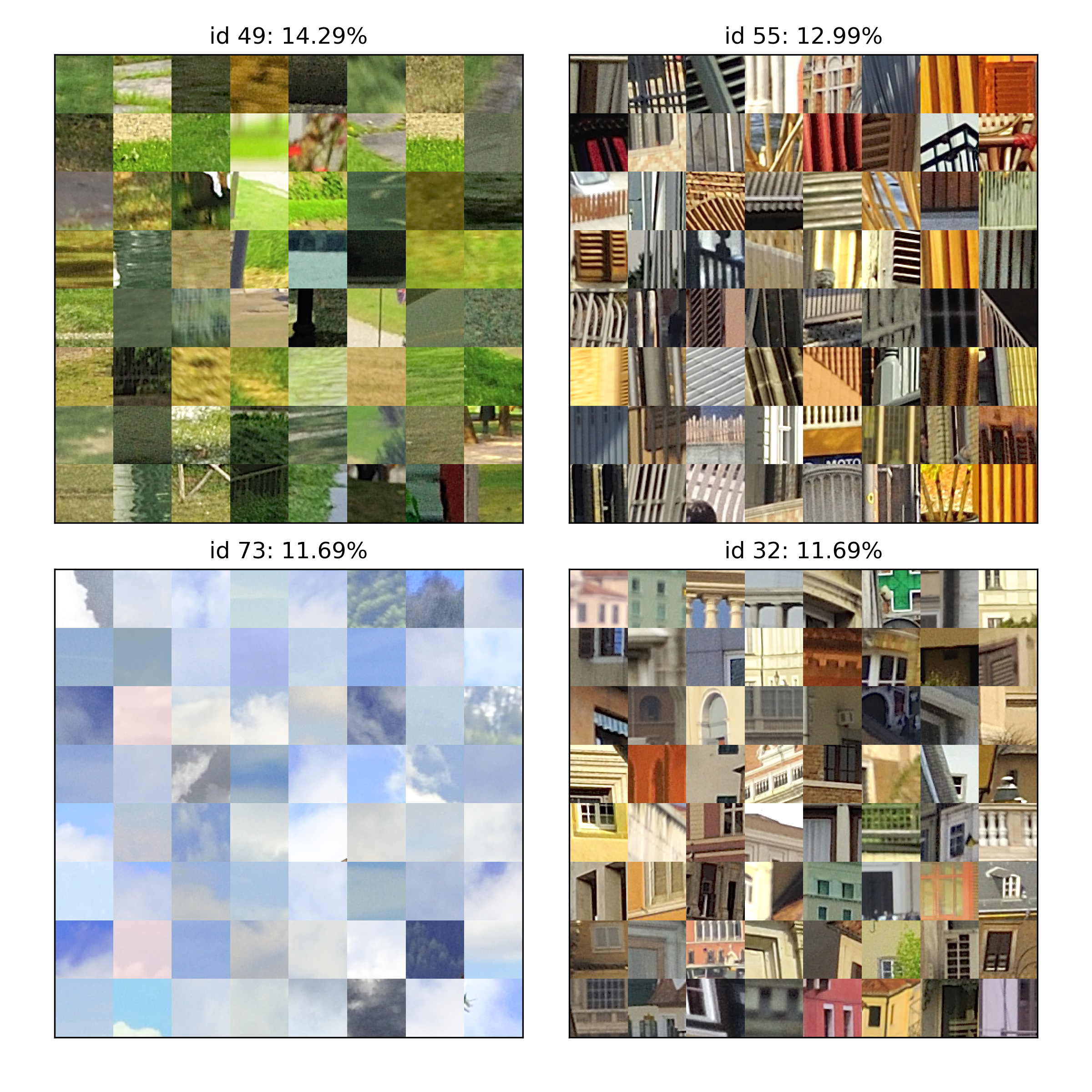} } \label{lh_texture} }%
\caption{A texture distribution example. (a) The lighthouse image, and (b) The top-4 textures of the lighthouse image. }%
\label{fig:kcd-texture}%
\end{figure}

We use the VGG-16 pre-trained \cite{simonyan2014very} network to extract bottleneck features from texture patches, then apply principal component analysis (PCA) to further reduce the dimension to 500, covering 82\% of the variance. Because we don't know how many types of textures in our training set, the K-Means algorithm is used to cluster textures into $K$ categories. On the RAISE dataset, we make a natural guess to select $K=100$. We argue that the number of classes is not essential. Still, just a dimension of our table pools, since different texture type's optimal tables will be aggregated in the prediction flow. The linear combination of optimal tables yields the final custom-tailored quantization table. Figure \ref{lh_texture} shows one example of texture mosaic images that composite the lighthouse (kodim19) image of the Kodak PhotoCD dataset. 

\SubSection{Annealing on Texture Mosaic Images} \label{sec_anneal}

The quantization table is an $8\times8$ matrix of 8-bit unsigned values. It's unrealistic to enumerate the whole space for an optimal solution. The quality degradation caused by quantization remains unknown so that we do not have an efficient approach like gradient descent to iterate the sample space. Meanwhile, it's likely to be trapped in local minimum during the optimization process. Therefore, we anneal the JPEG default luminance quantization table on the texture mosaic images with quality metric $Q=$50 and 95 to validate our approach. In each step, we randomly choose some table indices based on the magnitude of the current value. We weight smaller values more, assuming that smaller value has higher visual importance. Then we update the table values with $\pm1$ in probability to obtain a candidate solution. To be compliant with \texttt{libjpeg}, we scale the candidate solution to target quality metric using equation \eqref{eq:scale_quant}, where $S_f=1.0$ for $Q=50$ and $S_f=0.1$ for $Q=95$. We compress a new JPEG file using the candidate table as the luminance table and standard chrominance table. If the new JPEG file has size reduction and the quality degradation falls within tolerance $\gamma=0.01$, we accept the candidate solution and complete the current iteration. We evaluate the FSIM quality reduction by, 

\begin{equation}
\text{FSIM}(I_r, I_c) \ge \text{FSIM}(I_r, I_s) \times (1-\gamma), \label{eq:tolerance}
\end{equation}

where $I_r$ is the raw image, $I_c$ and $I_s$ is the JPEG image compressed with the candidate table and standard table. In order not to be trapped in a local minimum, there's a probability $P(i)$ to accept a worse solution, affected by the temperature function $T(i)$ and the energy delta $\Delta E$. The probability $P(i)$ to receive an answer is calculated by, 

\begin{equation}
P(i) = \Delta E \times T(i)\quad,\quad \text{for } i = 1\dots M
\end{equation}

\begin{equation}
\Delta E = \frac{S_i}{S_{i-1}}\quad,\quad S_i = C_i \times (1-D_i)\quad,\quad D_i=\text{FSIM}(I_r,I_c)
\end{equation}

\begin{equation}
T(i) = \frac{M}{M + i\times p},
\end{equation}

where $i$ is the iteration index, $M$ is the maximum iterations to anneal, $C_i$ denotes the current compressed JPEG file size, and $D_i$ is the FSIM quality distortion. The temperature function is designed so that the probability approaches $\frac{1}{p+1}$ at the end of the annealing process. With the design of probability $P(i)$, we have a higher possibility to accept a worse solution at the early stage, which prevents us from being trapped in a local minimum. And the probability decreases gradually with the number of iterations we run; then the annealing becomes a hill-climbing process. The current iteration could be finished by accepting a worse solution, or we'll randomly update again to get the next candidate. In this work, we choose $M=2,000$ and $p=10$ to anneal each texture mosaic image, taking around 5.5 hours to execute on an Intel Core i7-9700K processor core. 

\SubSection{Texture Training and Prediction} \label{sec_train}

After we cluster the texture mosaic images, we use the clustering labels to train a texture prediction model by supervised learning. With the ImageNet pre-trained VGG-16 network, we freeze the ConvNets parameters and fine-tune the fully connected layers as the classifier for 100 textures. We crop 261,712 texture patches from the RAISE dataset, split into 80\%-20\% for training and testing. We employ the Adam optimizer with default settings in PyTorch to fine-tune our network with learning rate 0.0001 and batch size 2,048. We train the texture CNN model for 30 epochs, resulting in top-3 testing accuracy around 99.77\%. For prediction, we crop the input image into $64\times64$ patches, then execute the forward propagation process to obtain each patch's texture category, forming a texture distribution to describe the image structure. 

\SubSection{Quantization Table Fusion}

It is common to see a very different distribution of textures, as pictures with all kinds of variety. To aggregate those per texture optimized quantization tables to better fit the whole image, we considered two strategies, voting by majority and weighted average. We select the weighted average policy for its overall better performance. The fused optimal quantization table $T_O$ is calculated by, 

\begin{equation}
T_O(i) = \sum_t ( T_t(i) \times W_t  )\quad,\quad \text{for } i = 0 \dots 63    \end{equation}


where $T_t$ denotes the optimal table of the texture $t$, and $W_t$ is the weights of the corresponding textures. The fused optimal quantization table of the lighthouse image is provided in Figure \ref{fig:fuse_lh}. Compared to the standard table, we mark the cells in red for increases and blue decreases. Our observation again validates the conclusions obtained from Monro and Sherlock \cite{monro1993optimum,sherlock1994model} that the JPEG standard table improperly over-estimates the low frequency parts and under-estimates the high frequencies. 

\Section{Experimental Results} \label{sec_expr}

We briefly describe the two datasets used in our experiments as follows: 

\textbf{RAISE:} The RAISE dataset \cite{dang2015raise} is a real-world camera photo database, collected from four photographers, capturing different scenes in over 80 places with different cameras. It consists of 8,156 high-resolution (around $4,288 \times 2,848$) RAW images. 

\textbf{Kodak:} The Kodak dataset \cite{kodakcd} has 24 lossless images, commonly used for evaluating image compression. Each image is about $768 \times 512$ in resolution. 

We choose the RAISE-1k subset of the RAISE dataset and randomly select 50 images as testing set to evaluate the optimality of JPEG compression. The rest of the images are used as the training set. Each image is cropped into $64\times 64$ patches with stride 256, generating total 261,712 texture patches. Then we cluster patches to 100 textures and perform the simulated annealing process on the stitched mosaic images as described earlier. At most 225 patches are randomly selected from each texture category to limit the required annealing time. To be fair, all the raw images are encoded using the JPEG standard table, the best table from Hopkins's work \cite{hopkins2018simulated}, and our fused quantization table, scaled to targeting JPEG quality as equation \eqref{eq:scale_quant}. The compression is done with the command line program \texttt{cjpeg} from \texttt{libjpeg} \cite{independent2016libjpeg}. Full reference image quality metric PSNR, SSIM, and FSIM are used as the quality benchmark, reported as the distance between the original image and the compressed JPEG image. 

\SubSection{RAISE Training Texture Annealing Result}

We present the annealing performance of the 100 texture mosaic images from the RAISE training patches at quality $Q=95$ in Table \ref{tab:perf-texture}. The proposed annealing method delivers slightly worse FSIM quality than Hopkins18, decreasing by 0.04\%, but it further reducing the JPEG size by 7.25\% on average. It is interesting to note that our method improves PSNR and SSIM by 1.38\% and 0.09\%. Although it is well known that PSNR does not align with the HVS, but it reflects the signal fidelity. We don't mind having higher PSNR if the compressed size is smaller. It somehow indicates that the optimized quantization table better adapts to the image content than standard and Hopkins' global table. 

\begin{table}[ht]
\caption{The RAISE training textures optimization performance, compared to Hopkins' global table at $Q=95$. We mark the superior results in bold. }
\label{tab:perf-texture}
\resizebox{\textwidth}{!}{%
\begin{tabular}{|l|l|l|l|l|l|l|l|l|}
\hline
\multirow{2}{*}{Q=95} & \multicolumn{4}{c|}{Hopkins18 v.s. Standard} & \multicolumn{4}{c|}{JQF v.s. Hopkins18} \\ \cline{2-9} 
 & Size & PSNR & SSIM & FSIM & Size & PSNR & SSIM & FSIM \\ \hline
100 Textures & \multicolumn{1}{r|}{-17.86\%} & \multicolumn{1}{r|}{-4.96\%} & \multicolumn{1}{r|}{-0.56\%} & \multicolumn{1}{r|}{-0.40\%} & \multicolumn{1}{r|}{\textbf{-7.25\%}} & \multicolumn{1}{r|}{\textbf{1.38\%}} & \multicolumn{1}{r|}{ \textbf{0.09\%}} & \multicolumn{1}{r|}{-0.04\%} \\ \hline
\end{tabular}%
}
\end{table}

\SubSection{Evaluation Result and Cross-validation} \label{sec_eval_raise}

The evaluation result of real-world images is reported in Table \ref{tab:perf-evaluate}. A similar pattern of better PSNR, SSIM and FSIM quality against Hopkins18 is also observed. The proposed JQF method achieves further 8.46\% size reduction and improve FSIM by 0.05\% on RAISE, while reduces 7.62\% compressed size and enhances FSIM quality by 0.09\% cross-validated on the Kodak dataset. Although we see a better Quality vs. Size trade-off compared to Hopkins' work, it can only be thought of as one possible outcome of the rate-distortion optimization process. If we compare the JQF annealing tables with the standard JPEG table, the 25.14\%-23.5\% compression gain does hurt the image quality, reducing the FSIM score by 0.28\%-0.35\%. Our experiences show that if we anneal fewer iterations or relax the tolerance $\gamma$ in equation \eqref{eq:tolerance}, we'll obtain optimal texture tables with better quality but worse compression ratio. As the storage cost becomes cheaper and cheaper, we should focus on the customized optimality, not the total compressed size. Therefore, we'll only compare the standard table when we scale the fused table at a different quality level. 

\begin{table}[ht]
\caption{The RAISE evaluation performance and cross-validation result on the Kodak dataset, compared to Hopkins' global table at $Q=95$. }
\label{tab:perf-evaluate}
\resizebox{\textwidth}{!}{%
\begin{tabular}{|l|r|r|r|r|r|r|r|r|}
\hline
\multirow{2}{*}{Q=95} & \multicolumn{4}{c|}{JQF v.s. Hopkins18} & \multicolumn{4}{c|}{JQF v.s. Standard} \\ \cline{2-9} 
 & \multicolumn{1}{l|}{Size} & \multicolumn{1}{l|}{PSNR} & \multicolumn{1}{l|}{SSIM} & \multicolumn{1}{l|}{FSIM} & \multicolumn{1}{l|}{Size} & \multicolumn{1}{l|}{PSNR} & \multicolumn{1}{l|}{SSIM} & \multicolumn{1}{l|}{FSIM} \\ \hline
RAISE testing set & \textbf{-8.46\%} & \textbf{1.96\%} & \textbf{0.22\%} & \textbf{0.05\%} & -25.14\% & -3.10\% & -0.52\% & -0.28\% \\ \hline
Kodak dataset & \textbf{-7.62\%} & \textbf{2.51\%} & \textbf{0.23\%} & \textbf{0.09\%} & -23.50\% & -2.90\% & -0.45\% & -0.35\% \\ \hline
\end{tabular}%
}
\end{table}

\SubSection{Scaling at Different Quality Metric}

As both Hopkins \cite{hopkins2018simulated} and Jiang \cite{jiang2011jpeg} report their results at different quality levels, it is clear that the default scaling equation \eqref{eq:scale_quant}, which uniformly scales quantization values is not a proper way to adapt quantization table to target quality. From Table \ref{tab:diff-q}, our experiments validate that our annealed optimal tables at $Q=95$ is only optimal at the trained level, i.e., the tables scaled to $Q=50$ have much worse Quality vs. Size trade-off than directly annealed at $Q=50$ (the right hand side). Since annealing texture images at all different qualities are not practical, we decide to use the annealed optimal tables at $Q=50$ and scale to other different qualities as our proposed solution. The optimal tables from $Q=50$ maintain good Quality vs. Size reduction trade-off at different quality levels. 

\begin{table}[ht]
\caption{The Kodak performance impact of annealed tables scaled to different $Q$}
\label{tab:diff-q}
\resizebox{\textwidth}{!}{%
\begin{tabular}{|l|r|r|r|r|l|r|r|r|r|}
\hline
\multirow{2}{*}{Anneal Q=95} & \multicolumn{4}{c|}{JQF v.s. Standard} & \multirow{2}{*}{Anneal Q=50} & \multicolumn{4}{c|}{JQF v.s. Standard} \\ \cline{2-5} \cline{7-10} 
 & \multicolumn{1}{l|}{Size} & \multicolumn{1}{l|}{PSNR} & \multicolumn{1}{l|}{SSIM} & \multicolumn{1}{l|}{FSIM} &  & \multicolumn{1}{l|}{Size} & \multicolumn{1}{l|}{PSNR} & \multicolumn{1}{l|}{SSIM} & \multicolumn{1}{l|}{FSIM} \\ \hline
Scale to 35 & -41.73\% & -9.30\% & -10.22\% & -2.36\% & Scale to 35 & -24.19\% & -3.91\% & -4.13\% & -1.01\% \\ \hline
Scale to 50 & -38.67\% & -7.99\% & -7.24\% & -1.82\% & Scale to 50 & -21.65\% & -3.06\% & -2.71\% & -0.73\% \\ \hline
Scale to 75 & -32.57\% & -5.69\% & -3.30\% & -0.98\% & Scale to 75 & -17.49\% & -1.71\% & -1.05\% & -0.38\% \\ \hline
Scale to 95 & -23.50\% & -2.90\% & -0.45\% & -0.35\% & Scale to 95 & -12.34\% & -0.58\% & -0.09\% & -0.09\% \\ \hline
\end{tabular}%
}
\end{table}

\SubSection{Prediction Computational Cost}

The image specific optimization only matters if the extra computational cost is within an acceptable range. For the texture prediction, we don't need to predict the full resolution of the given image, but on a down-sampled version of it, say around $2,048 \times 1,360$ is enough. We used a workstation with Intel Core i7-9700K CPU and Nvidia GeForce RTX 2080 Ti GPU to complete our experiments, taking about 0.48 seconds with GPU and 23.83 seconds on pure CPU computation per RAISE testing image, as shown in Table \ref{tab:prediction-time}.  

\begin{table}[ht]
\centering
\caption{The average texture prediction time, in seconds. }
\label{tab:prediction-time}
\resizebox{0.6\textwidth}{!}{%
\begin{tabular}{|l|r|r|r|}
\hline
Database & \multicolumn{1}{l|}{\# Images} & \multicolumn{1}{l|}{GPU time} & \multicolumn{1}{l|}{CPU time} \\ \hline
RAISE testing set & 50 & 0.4839 & 23.8342 \\ \hline
Kodak dataset & 24 & 0.0438 & 2.6862 \\ \hline
\end{tabular}
}
\end{table}

Prediction on a smaller resized image could further reduce the time needed for prediction, but it may not be necessary. In fact, in our experience, we found that the  computational time of image quality metric FSIM is a crucial factor that limits the optimization process. Even so, the annealing and CNN texture training process can be implemented offline without affecting our proposed JQF as a real-time JPEG optimization approach. 

\Section{Conclusion}

We propose a novel JPEG Quantization Table Fusion method using simulated annealing on texture mosaic images to search for an optimal quantization table for each texture category. A VGG-16 pre-trained CNN model is used to learn those texture features and predict the input image's texture distribution, then fuse optimal texture tables to come out an image specific optimal quantization table. Our method shows superior performance on a larger consumer photo dataset and generalizes well on cross-database evaluation. On the Kodak dataset with quality $Q=95$ setting, our experiment shows a size reduction of 23.5\% over the JPEG standard table with a slightly 0.35\% FSIM decrease but visually unperceivable. The per-image optimality for JPEG encoding is achieved with less than one second additional timing cost. 

\Section{References}
\bibliographystyle{IEEEbib}
\bibliography{refs}

\end{document}